\documentclass{amsart}

\usepackage[utf8]{inputenc} 
\usepackage[T1]{fontenc}    
\usepackage{lmodern}        
\usepackage{hyperref}       
\usepackage{url}            
\usepackage{booktabs}       
\usepackage{amsfonts}       
\usepackage{nicefrac}       
\usepackage{microtype}      
\usepackage{graphicx}
\usepackage{mymacros,todonotes}
\usepackage{dsfont}
\usepackage{algorithm}
\usepackage{algpseudocode}
\usepackage{tikz,xcolor}
\usepackage{comment}
\usepackage{chngcntr}
\counterwithin{figure}{section}

\definecolor{darkgreen}{rgb}{0.0, 0.3, 0.0}
\definecolor{darkred}{rgb}{ 0.6,0, 0.0}
\definecolor{darkblue}{rgb}{0,0,0.5}

\usepackage{color}
\usepackage{fancyvrb}

\DefineVerbatimEnvironment{Highlighting}{Verbatim}{commandchars=\\\{\}}
\usepackage{framed}
\definecolor{shadecolor}{RGB}{248,248,248}

\usepackage{longtable,booktabs,array}
\usepackage{calc} 
\usepackage{etoolbox}
\makeatletter
\patchcmd\longtable{\par}{\if@noskipsec\mbox{}\fi\par}{}{}
\makeatother
\IfFileExists{footnotehyper.sty}{\usepackage{footnotehyper}}{\usepackage{footnote}}
\makesavenoteenv{longtable}
\usepackage{calc} 

\usepackage[round]{natbib}

\begin{document}

\newcommand{\sgn}{\operatorname{sgn}}
\newcommand{\SI}{\mathcal{S}}
\newcommand{\XS}{X^{\mathrm{S}}}
\newcommand{\XD}{X^{\mathrm{D}}}
\newcommand{\pS}{p^{\mathrm{S}}}
\newcommand{\pD}{p^{\mathrm{D}}}
\nc{\hpsi}{\widehat{\Phi}}
\newcommand{\J}{\mathcal{J}}

\title{The R\'enyi Outlier Test}

\author{Ryan Christ}
\author{Ira Hall}
\author{David Steinsaltz}

\address[Ryan Christ and Ira Hall]
{Center for Genomic Health \& Department of Genetics\\
	Yale University School of Medicine\\
	New Haven, CT USA}
\email[Ryan Christ]{ryan.christ@yale.edu}
\email[Ira Hall]{ira.hall@yale.edu}

\address[David Steinsaltz]
{Department of Statistics\\
	Oxford University\\
	Oxford, UK}
\email[David Steinsaltz]{steinsal@stats.ox.ac.uk}


\begin{abstract}
Cox and Kartsonaki proposed a simple outlier test for a vector of p-values based on the {\em R\'enyi transformation} that is fast for large $p$ and numerically stable for very small p-values -- key properties for large data analysis. We propose and implement a generalization of this procedure we call the R\'enyi Outlier Test (ROT). This procedure maintains the key properties of the original but is much more robust to uncertainty in the number of outliers expected \textit{a priori} among the p-values. The ROT can also account for two types of prior information that are common in modern data analysis. The first is the prior probability that a given p-value may be outlying. The second is an estimate of how far of an outlier a p-value might be, conditional on it being an outlier; in other words, an estimate of effect size. Using a series of pre-calculated spline functions, we provide a fast and numerically stable implementation of the ROT in our R package \texttt{renyi}.
\end{abstract}

\maketitle

\newcommand{\Hg}{H_{\mathrm{g}}}
\newcommand{\Ho}{\mathcal{H}}

Cox and Kartsonaki proposed an outlier test based on the {\em R\'enyi transformation} $\rho : [0,1]^p \rightarrow [0,\infty]^p$ \cite{cox2019analysis}. 	For an ordered vector $u \in [0,1]^p$, such that $u_1 \leq u_2 \leq \ldots \leq u_p$, $\rho(u)_j = j \log\left(u_{j+1} \big/ u_j\right)$ for all $j < p$ and $\rho(u)_p =  -p\log\left(u_p\right)$.
Alfr\'ed R\'enyi pointed out that when $\rho$ is applied to a vector $U$ of ordered independent uniform random variables, the image $\rho\left(U\right)$ has entries that are independent exponential random variables \cite{AR53}. Building on this observation, for some user-specified number of potential outliers, $k$, Cox and Kartsonaki proposed testing the null hypothesis $H_0$ that an observed $u$ is a vector of ordered independent uniform random variables by comparing $\sum\limits_{j=1}^k \rho\left(u\right)_{j}$ against its null distribution: a Gamma distribution with shape $k$ and rate 1. This simple procedure allows for the rapid calculation of numerically precise p-values even when $p$ is very large and the p-value returned is in the lower ranges accessible to machine precision. However, the power depends sensitively on  the {\it a priori} specification of the number of outliers $k$. 
A more robust method would maintain power in the more common situation where the number $k$ of outliers is unknown, but it is possible to specify a rough upper bound $K$ on the likely number of outliers. 

We present a robust generalization of Cox and Kartsonaki's proposal that only requires an approximate upper bound $K$. Our generalization also admits two types of prior information that is common in modern applications can be used to sharpen the alternative hypothesis and thereby improve power. The first, $\pi \in \mathbb{R}_{\geq 0}^{p}$, is taken to be proportional to the prior probability that a given uniform random variable is an outlier. The second, $\eta \in \mathbb{R}_{\geq 0}^{p}$, is related to effect size: how far outlying $u_j$ will be given that it is an outlier. In the common context where each element of $u$ can be thought of as a p-value for testing whether some coefficient $\beta$ in a linear regression model is zero, we take $\eta_j \propto \mathbb{E}\left[\left. \beta_j^2 \right| \beta_j \neq 0\right]$. In the absence of prior information or expectations, we take the neutral defaults $\pi_j = 1$ and $\eta_j = 1$ for all $j$.  Critically, our approach, which we call the R\'enyi Outlier Test (ROT), maintains the computational speed and numerical precision of the original test proposed by Cox and Kartsonaki. We also provide the \texttt{renyi} R package that implements our procedure, making use of pre-calculated spline functions. The package is publicly available at \url{ryanchrist.r-universe.dev/renyi}.

Compared to more commonly used ``minimum''-based approaches, such as testing the minimum p-value with Bonferroni correction or Holm's method,
Higher Criticism and related tests in the General Goodness of Fit test family have more power when, roughly speaking, there are a handful of modestly small p-values \cite{donoho2004higher, hz19, zhang2022general}. Computational speed and numerical precision have been major obstacles to applying these outlier tests in practice. Recently, Wang et al. proposed a fast implementation of higher criticism that is numerically stable for even very small p-values \cite{wang2024accurate}. However, this approach does not admit prior information such as $\pi$ and $\eta$. 

Given an initial set of \textit{unordered} uniform random variables $U \in [0,1]^p$, $\pi$, and $\eta$, the ROT is a two step procedure to test the null hypothesis $H_0 : U_j \stackrel{iid}{\sim} \mathrm{Unif}(0,1)$ for $j = 1,\ldots, p$. Note that if each $U_j$ represents a p-value, they must be {\bf exactly uniform} under the global null hypothesis, not sub-uniform or super-uniform. First, we perform a simple generalization of the \textit{R\'enyi transformation} which accounts for $\pi$ and $\eta$ to obtain a set of independent standard exponential random variables. We then test the outliers based on those exponential random variables using a procedure robustified to our choice of $K$. 

Define 
\begin{equation}
Z_j = \eta_j \left(-\log\left(U_j\right) + \log\left(\pi_j\right)\right)  = \eta_j \left(-\log\left(U_j\right) \right) + \zeta_j
\end{equation}
where $\zeta_j = \eta_j\log\left(\pi_j\right)$, and let $N: \mathbb{R} \to \mathbb{N}$ be the corresponding point process:
\begin{equation}
    \label{E:pointprocess}
    N(t) := \sum_{j=1}^p \mathbf{1}\{Z_j \le t \} .
\end{equation}
For $-\infty\le t < \infty$, define a filtration by letting $\mathcal{F}_t$ be the sigma algebra generated by all events of the form 
$$
    \bigl\{-\log(U_j)\le \frac{s}{\eta_j} - \log\pi_j \bigr\}
$$
for $s\le t$ and $1\le j \le p$.
We understand $\eta_j$ and $\pi_j$ to be measurable with respect to $\mathcal{F}_t$ for all $t$ (including $t=-\infty$).
$N(t)$ is adapted with respect to this filtration, and the compensator is
\begin{equation}
    \label{E:compensator}
    \Lambda(t):= \sum_{j=1}^p \eta_j^{-1} \bigl( t\wedge Z_j - t \wedge \zeta_j).
\end{equation}
Since $\Lambda: \mathbb{R} \to [0, -\sum \log U_j]$ is a continuous non-decreasing function, it has a right-inverse $\Lambda^{-1}: [0,-\sum\log U_j] \to [\min\zeta_j, \max Z_j]$ defined by $\Lambda^{-1}(u) = \sup\{t: \Lambda(t)<u\}$; that is, $\Lambda \circ \Lambda^{-1}$ is the identity map on $[0,-\sum\log U_j]$.
Then by Theorem 15.15 of \cite{oK21} $N\circ \Lambda^{-1}$ is a Poisson process with unit rate (up to the time of the $p$-th event).
The test will then be based on the statistics $(X_1,\dots,X_p)$, which are the interarrival times of the process, in reverse order; under the global null hypothesis these are i.i.d.\ unit exponential random variables.
This is a generalization of the original \emph{R\'enyi transformation}. 
 
Let $G_{x}$ denote the CDF of the Gamma distribution with shape parameter $x$ and rate 1, and let $I_{x,y}$ denote the CDF of the Beta distribution with mean $x/(x+y)$. If the number of outliers $k$ were known, then the p-value $1- G_k\left(\sum\limits_{j=1}^k X_j\right)$ would provide a  well-powered test of $H_0$ against alternatives where $k$ of the original values $U_j$ are sampled from a distribution that makes them substantially smaller than uniform [0,1] random variables.
If $k$ is chosen too small then some potential power is lost, while a too-large $k$ would crush the power by mixing the true outliers with non-outlying observations. To mitigate this weakness, then, we try to find an upper bound $K$ on $k$, and perform an omnibus test over different $k$ up to $K^\star := 2^{\lceil \log_2 {K} \rceil }$. Let $\wt{X}_j = X_j$ for all $j< K^\star$ and define
\begin{equation} \label{E:XKstar}
\wt{X}_{K^\star} = -\log\Bigl(1-I_{p-K^\star+1,K^\star}\Bigl( \exp\Bigl(-\sum\limits_{j=K^\star}^p \frac{X_{j}}{j}  \Bigr)\Bigr)\Bigr).
\end{equation}
Using R\'enyi's representation for the order statistics of independent exponential random variables, the sum in \eqref{E:XKstar} is distributed (under the global null hypothesis) as the $K^*$-th largest out of $p$ independent unit exponential random variables \cite{AR53}.
Exponentiating this as above yields the corresponding order statistic of Uniform random variables, which can be transformed by the Beta cdf to a new Uniform random variable, leaving us at last with $\wt{X}_{K^*}$ being another unit exponential random variable, independent of $\wt{X}_1,\dots,\wt{X}_{K^*-1}$.

We now define the ROT test statistic as
\begin{equation} \label{eq:rot}
	\rho_{K^\star}= \max \limits_{i \in \mathcal{I}_k} \ -\log\left( 1-G_i\left( \sum\limits_{j=1}^i \wt{X}_j  \right) \right)
\end{equation}
where
$\mathcal{I}_k = \left(1,2,4,8,\ldots, K^\star\right)$. The fact that each $\wt{X}_j$ in \eqref{eq:rot} is an independent exponential, makes simulating the null distribution of $\rho_{K^\star}$ straightforward. We used Monte Carlo simulation to estimate the body of the null distribution of $\rho_{K^\star}$ for $K^\star$ taking values in $\left(1,2,4,\ldots,128\right)$. We used those null simulations to fit a line to the log-linear tail of each distribtion and fit a cubic spline function to the body of the distribution. This yielded a compressed form of a lookup table for each test statistic that allows rapid computation p-values for a wide range of $K^\star$. It is available via our R package \texttt{renyi}. The package is publicly available at \url{ryanchrist.r-universe.dev/renyi}.

\newpage
\bibliographystyle{plainnat}
\bibliography{renyi.bib}

\end{document}